\documentclass{optica-article}
\usepackage{newunicodechar}
\usepackage{textcomp}
\newunicodechar{ }{\,}
\journal{opticajournal} 

\articletype{Research Article}

\usepackage{graphicx} 
\usepackage{caption}
\usepackage{subcaption}
\usepackage{hyperref}
\usepackage[normalem]{ulem}
\usepackage{lineno} 
\hypersetup{
  colorlinks = true,
  linkcolor = blue
}

\usepackage{eso-pic}
\AddToShipoutPictureFG*{
  \AtPageLowerLeft{%
    \raisebox{\dimexpr+0.65in}{\hspace*{0.7in}%
      \scalebox{1}{\color{gray}
        \parbox{\textwidth}{\footnotesize 
         This manuscript has been authored by UT-Battelle, LLC, under contract DE-AC05-00OR22725 with the US Department of Energy (DOE). The US government retains and the publisher, by accepting the work for publication, acknowledges that the US government retains a non-exclusive, paid-up, irrevocable, world-wide license to publish or reproduce the submitted manuscript version of this work, or allow others to do so, for US government purposes. DOE will provide public access to these results of federally sponsored research in accordance with the DOE Public Access Plan (\url{https://www.energy.gov/doe-public-access-plan}).
        }%
      }%
    }%
  }%
}%

\begin{document}

\title{A Robust COTS Objective for Diffraction-Limited, High-NA, Long Front Working Distance Imaging}

\author{Jiafeng Cui,\authormark{1,2,*} Gilles Buchs,\authormark{1,2} and Christopher M. Seck\authormark{1,2,\textdagger}}

\address{
\authormark{1}Quantum Information Science Section, Computational Sciences and Engineering Division, Oak Ridge National Laboratory, Oak Ridge, TN\\
\authormark{2}Quantum Science Center, Oak Ridge National Laboratory, Oak Ridge, TN, USA\\
}

\email{\authormark{*}cuij@ornl.gov}
\email{\authormark{\textdagger}seckcm@ornl.gov} 


\begin{abstract*} 
We present a robust objective lens optimized for applications requiring both high numerical aperture (NA) and long front working distance imaging comprised of all commercial-off-the-shelf (COTS) singlet lenses. Unlike traditional designs that require separate collimation and refocusing stages, our approach directly converges imaged light to the back focal plane using a single lens group. Our configuration corrects spherical aberrations and efficiently collects light to achieve diffraction-limited performance across a wide range of wavelengths while simplifying alignment and assembly. Using this approach, we design and construct an example objective lens that features a long front working distance of 61~mm and a clipped NA of 0.30 (limited by an aperture in our experimental setup). We experimentally verify that it achieves monochromatic diffraction-limited resolution at wavelengths from 375~nm to 866~nm without requiring replacement of the lenses or changing the inter-lens spacings, and its performance remains robust across a 46~mm range variation in total length (by adjusting mainly the back working distances). Additionally, we develop a quantitative method to measure the field of view (FOV) using an experimentally-calibrated pinhole target. Under 397~nm illumination (i.e. from $^{40}$Ca$^+$ ion fluorescence), the objective achieves a resolution of 0.87~$\mu$m with a 540~$\mu$m FOV. This robust, all-COTS, and versatile design is well-suited for a broad range of experiments, supporting high-precision measurements and exploring quantum phenomena.

\end{abstract*}

\section{Introduction}
Optical detection and readout play crucial roles in quantum system studies, whether in capturing state-dependent fluorescence emitted by trapped neutral atoms, molecules\cite{Bakr2009QuantumGas}, and/or ions\cite{Blatt2008Entangled}, or in collecting diffused light in optomechanical systems (e.g. levitated nanospheres) \cite{Millen2020Opomechanics} . Achieving site-resolved detection requires an optical imaging system with both a high numerical aperture (NA) for efficient light collection and diffraction-limited performance to resolve individual sites or particles. However, these experiments often involve object(s) of interest housed within a vacuum chamber, sometimes placed inside a cryogenic system, requiring imaging through multiple thick glass windows at long working distances \cite{Pagano2019Cryogenic}. These constraints introduce significant optical and mechanical challenges.

One imaging approach is to position an infinity-corrected objective within the vacuum chamber, including single concave mirror(s) \cite{Maiwald2012Collecting,Chou2017Single} or used in conjunction with a lens \cite{Araneda2020Panopticon,Carter2024Ion}, allowing for a short front working distance and enabling a high NA to maximize light collection. However, \textit{in vacuo} optics and any associated coatings must be vacuum-compatible, often bakeable at high temperatures, be mounted on a movable fixture that may limit optical access, and are often custom optics with associated high cost\cite{Coyne2014LIGO,Robens2017High,Carter2024InVacuum}.

A competing imaging approach utilizes an external imaging system, requiring objective designs with both long front working distances to ensure access to the object(s) of interest and a corresponding high NA. However, these requirements introduce significant spherical aberrations, which must be compensated for in the objective design. One route to accomplish this is to employ a customized aspherical lens, which is both costly and time consuming to design and fabricate\cite{Jechow2011Wavelengthscale}. Another approach is to design a compound objective lens, comprising of several singlet lenses with fixed inter-lens spacings, that can provide diffraction-limited performance, e.g. the design presented by Alt in Ref.~\cite{alt2002objective}. Similar designs using only commercial-off-the-shelf (COTS) components have been published \cite{Bennie13Versatile,Pritchard2016Long,Li2018High,Li2020High}. However, these designs require additional optics before the imaging apparatus to refocus the collimated image, often resulting in a bulky and sophisticated setup that may obstruct the optical path, therefore restricting the effective system NA. These limitations can restrict both the efficient and effective study of atomic, molecular, and optical physics (AMOP) and quantum information science (QIS) experiments including high precision and high accuracy measurements utilizing atomic or molecular neutrals and/or ions and/or small particles. One potential improvement over the design presented in Ref. \cite{alt2002objective} is to directly project the image onto the back focal plane of the compound objective, eliminating the need for extra refocusing optics. While this approach has been explored in design\cite{Wong-Campos2016HighResolution} and simulation\cite{silvan2020objective} studies, to the best of our knowledge, further experimental investigations appear to be limited.

In this manuscript, we present a robust, diffraction-limited imaging system leveraging readily available COTS components to achieve a high NA and long front working distance. It employs a compound objective lens configuration that is optimized for (i) the least number of lenses to correct for spherical aberrations and (ii) directly converges captured light to the back focal plane, which simplifies construction and alignment. Specifically, our example objective addresses challenges imposed by vacuum and/or cryogenic environment physical constraints, resulting in a long front working distance of 61~mm and a NA of 0.30, limited by an aperture in our experimental setup (the unclipped NA is 0.37). 

We experimentally verify that the objective achieves monochromatic diffraction-limited resolution at wavelengths from 375~nm to 866~nm without requiring replacement of the lenses or changing the inter-lens spacings, and its performance remains robust across a 46~mm range variation in total length (by mainly changing the back working distances with minimal impact on performance. Furthermore, we develop a method to quantitatively measure the field of view (FOV) using an experimentally calibrated target. When illuminated with 397~nm light (i.e. to image $^{40}$Ca$^+$ ion fluorescence), the objective reaches a resolution of 0.87~$\mu$m with a 540~$\mu$m FOV.

This manuscript is structured as follows: Section~\ref{Sec:Method} outlines the general design methodology for our objective design; Section~\ref{Sec:Design} details the design and construction of our specific example objective; Section~\ref{Sec:Diffraction} presents the experimental evaluation of its diffraction-limited performance; and Section~\ref{Sec:FOV} describes the novel FOV measurement method. The performance, flexibility, and versatility of this innovative COTS design make it applicable to a broad range of experiments, enabling the exploration of a diverse set of quantum phenomena and high-precision measurements.

\section{Methodology}\label{Sec:Method}

In AMO physics, most imaging systems require long front working distances with a high NA. Achieving both inevitably introduces optical aberrations that degrade image quality. The objective presented in this manuscript features a long front working distance and projects a direct, magnified image onto the back focal plane, while maintaining a high NA (e.g. the NA is established according to physical access constraints or design requirements) and diffraction-limited performance for efficient light collection and optimal resolution.

To accomplish this, we target an objective design with a compound configuration utilizing COTS lenses with precisely engineered spacing rings. The first element typically is a meniscus lens, which helps maintain the desired NA while minimizing spherical aberrations, similar to the collimated beam design in Ref.\cite{alt2002objective}. Subsequent plano-convex lenses converge the image onto the back focal plane.A plano-concave lens is used as the final element to compensate for aberrations introduced by the prior positive focal length lenses.

The design process, described below, employs an iterative approach using Ansys Zemax OpticStudio ray tracing software \cite{Bennie13Versatile, Pritchard2016Long, Li2018High, Li2020High, ZemaxUserManual}.
\begin{enumerate}
    \item The process begins by defining the monochromatic image target (atoms or particles with designated wavelength) and the placement of flat window(s) with specified thickness(s) based on the experimental setup. Constraints on front and back working distances, dictated by physical space limitations, as well as NA and magnification requirements, are incorporated throughout the process.
    
    \item Starting with a fundamental 3-lens configuration (meniscus, plano-convex, and plano-concave oriented from object to image), we initiate the iterative numerical optimization process. This involves adjusting the curvature of the lens(es), center thickness(es), and the air gaps between them. Because the ray-tracing optimization algorithm does not consider diffraction effects, the primary goal in this step is to minimize the root mean square (RMS) wavefront error, ensuring the focused spot size at the image plane remains below the Airy-disk radius. This will ensure that the real image will be diffraction-limited (Rayleigh criterion).
    
    \item The initial 3-lens design may not sufficiently correct higher-order optical aberrations to meet experimental demands. To compensate for this, we employ an iterative lens addition strategy. Additional lenses are inserted between the first meniscus and the final plano-concave lens. Following each lens addition, the numerical optimization process described in Step 2 is repeated. This iterative approach continues until the focused spot size achieves diffraction-limited performance.
    
    \item Each optical element is sequentially replaced with a closely-matching COTS singlet lens, and the remaining parameters are re-optimized after each substitution to preserve diffraction-limited performance.
    
    \item After all optimizations are complete, the objective is designed. All performance metrics can then be simulated and performance at other operating wavelengths simulated via optimizing the front working distance.
\end{enumerate}
This methodology ensures a robust, cost-effective, and high-performance optical design suitable for a wide range of experiments.

\section{Design and Construction of Example Objective}\label{Sec:Design}

\begin{figure}[ht]
  \centering
  \includegraphics[width = 10cm]{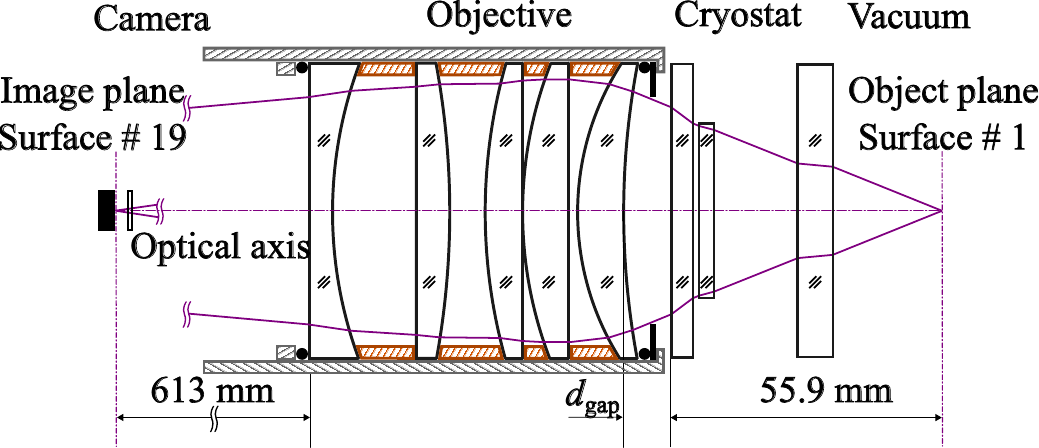}
  \caption{Cross-section of the example imaging objective. The system comprises (from the right to the left): vacuum viewport, two cryostat viewports, compound objective with 5 singlet lenses, and the camera system (including sensor window). Detailed parameters are provided in Table~\ref{tab:prescription}. Note that the NA is limited by a 3~mm thick, 30~mm diameter cryostat viewport flat window (Window 1 in Table~\ref{tab:prescription}). Custom-made brass spacer rings (orange) with a 2~inch outer diameter and 2~mm wall thickness are placed between the lenses and feature surface curvatures that match the lens contact surfaces. Their lengths are precisely calculated based on the lens-to-lens distances specified in Table~\ref{tab:prescription}. The solid black circles against the first and the last lenses are o-rings, both necessary to reduce mechanical stress on the lens stack. An anodized aluminium aperture mask, placed before the first lens, blocks stray light from outside the constrained numerical aperture. $d_\textrm{gap}$ is the distance from the last window to the first lens center distance.
  \label{fig:layout}
  }
\end{figure}

The example objective described below is designed for direct imaging of 397~nm light ($^{40}$Ca$^+$ ion $^2$S$_{1/2}$ to $^2$P$_{1/2}$ transition wavelength) onto the back focal plane with diffraction-limited performance. We limit the NA to 0.30 to simulate a cryogenic vacuum chamber viewport aperture (the unclipped NA is 0.37), as illustrated in Figure \ref{fig:layout}. This constraint on the NA, combined with readily available COTS components, led to choosing 2~inch diameter fused silica singlets for all lenses. 

\begin{table}
  \centering
  \caption{Example objective design using Ansys Zemax OpticStudio with fixed $\textrm{NA} = 0.30$. The subsystems are vacuum, cryostat, objective, and camera. $^1$ $(^2)$ denotes a ThorLabs (EKSMA Optics) part number. All have 2~inch (50.8~mm) diameter. Note that all lenses used in experiments utilize an anti-reflection (AR) coating optimized for 350~nm to 700~nm light.
  \label{tab:prescription}
  }
  \begin{tabular}{lccccl} 
\hline \hline
Surface & Curvature (mm) & Thickness (mm) & Material & Optic & System\\
\hline
1&  $\infty$&28.2&vacuum& &Vacuum\\ 
2&  $\infty$&6.4&silica&Viewport &\\

\hline
3&  $\infty$&13.7&vacuum& &Cryostat\\
4&  $\infty$&3.0&silica&Window 1&\\
5&  $\infty$&0.6&vacuum& &\\
6&  $\infty$&4.0&silica&Window 2&\\

7&  $\infty$&5.0&air& &\\
\hline
8&  135.3&7.8&silica&LE4125$^1$&Objective\\
9&  46.5&3&air& &\\
10&  $\infty$&7.8&silica&LA4904$^1$&\\
11&  69&0&air& &\\
12&  $\infty$&6.6&silica&LA4984$^1$&\\
13&  92&6&air& &\\
14&  -138&5.4&silica&LA4855$^1$&\\
15&  $\infty$&14.6&air& &\\
16&  71.4&3.2&silica&112-1512E$^2$&\\
17&  $\infty$&610&air& &\\

\hline
18&  $\infty$&1&Silica&Window &Camera\\
19&  $\infty$&2&Vacuum& &\\
\hline

\end{tabular}
\end{table}

The dominant aberrations in the presented system arise from the combination of the relatively large NA and the thick fused silica flat windows between the object (ion in our example case) and the first objective lens (total flat window thickness 13.4 mm). Following the method described in Section~\ref{Sec:Method} to achieve diffraction-limited performance while projecting a magnified (at least 10x) and focused ion image directly onto the back focal plane, the final objective configuration utilizes five 2~inch diameter COTS lenses along with precisely designed spacing rings (as shown in Figure~\ref{fig:layout}). All objective components are assembled within a standard 3~inch long, 2~inch diameter lens tube (Thorlabs SM2L30) (note that the external threads closest to the object are machined off). The precisely machined brass spacer rings are critical to the assembly. All match the surface curvatures of the lenses (machine length and radius error tolerances are 10 um) to (i) ensure proper lens separation, (ii) assist in aligning all lens optical centers, (iii) maximize the lens-to-ring contact area, and (iv) prevent lens-to-lens contact damage during assembly. We use a retention ring and two o-rings to secure the lenses and spacing rings without applying excessive pressure. The complete lens surfaces details such as curvature, thickness, material, manufacturer, and order are detailed in Table~\ref{tab:prescription}. The final objective design is in Ref.\cite{Zemaxfile}

\begin{figure}[ht]
   \centering
   \begin{subfigure}[b]{0.48\textwidth}
     \centering   
     \caption{\label{fig:strehl ratio}}
     \includegraphics[width=\textwidth]{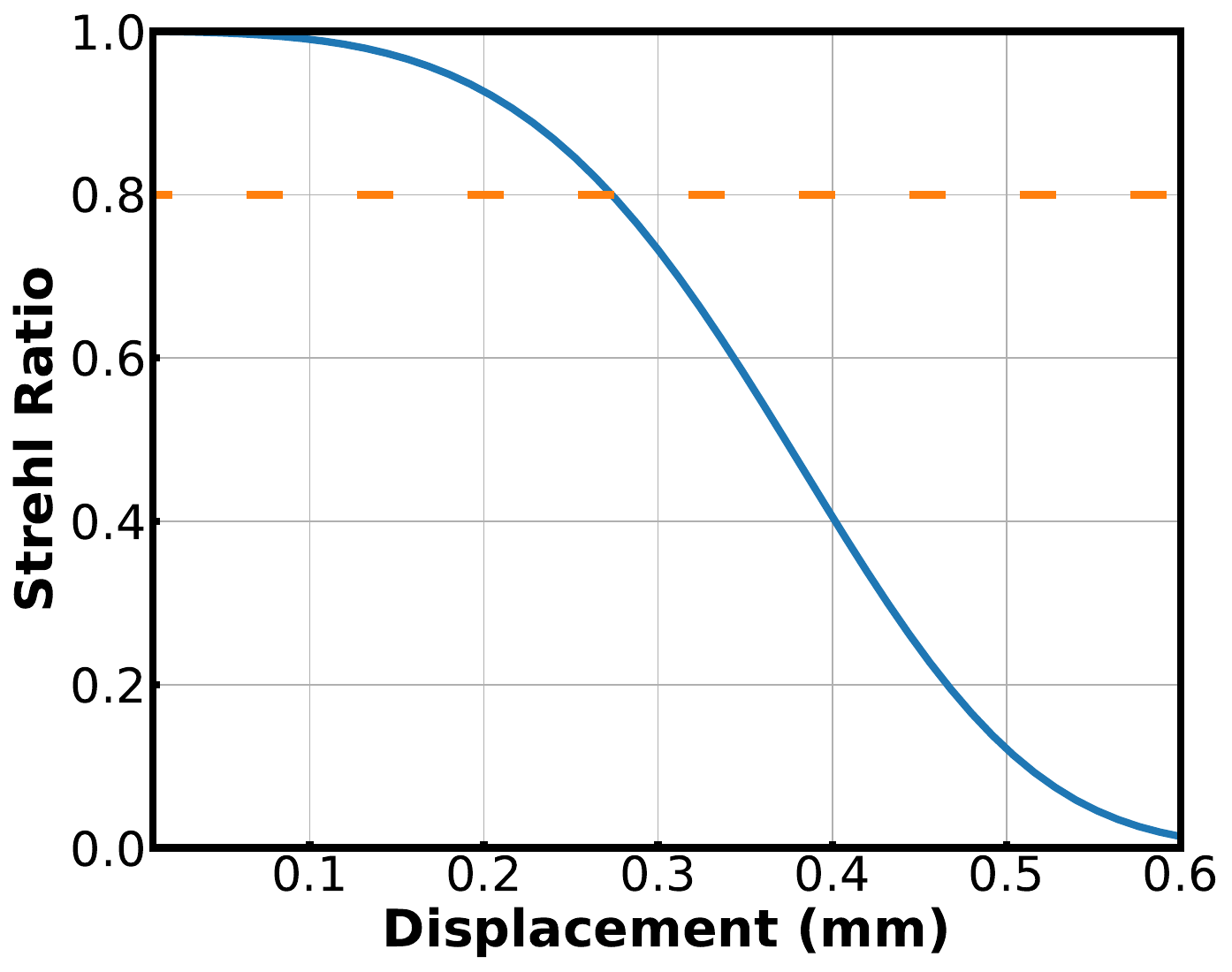}
   \end{subfigure}
   \begin{subfigure}[b]{0.48\textwidth}
     \centering
     \caption{\label{fig:MTF}}
     \includegraphics[width=\textwidth]{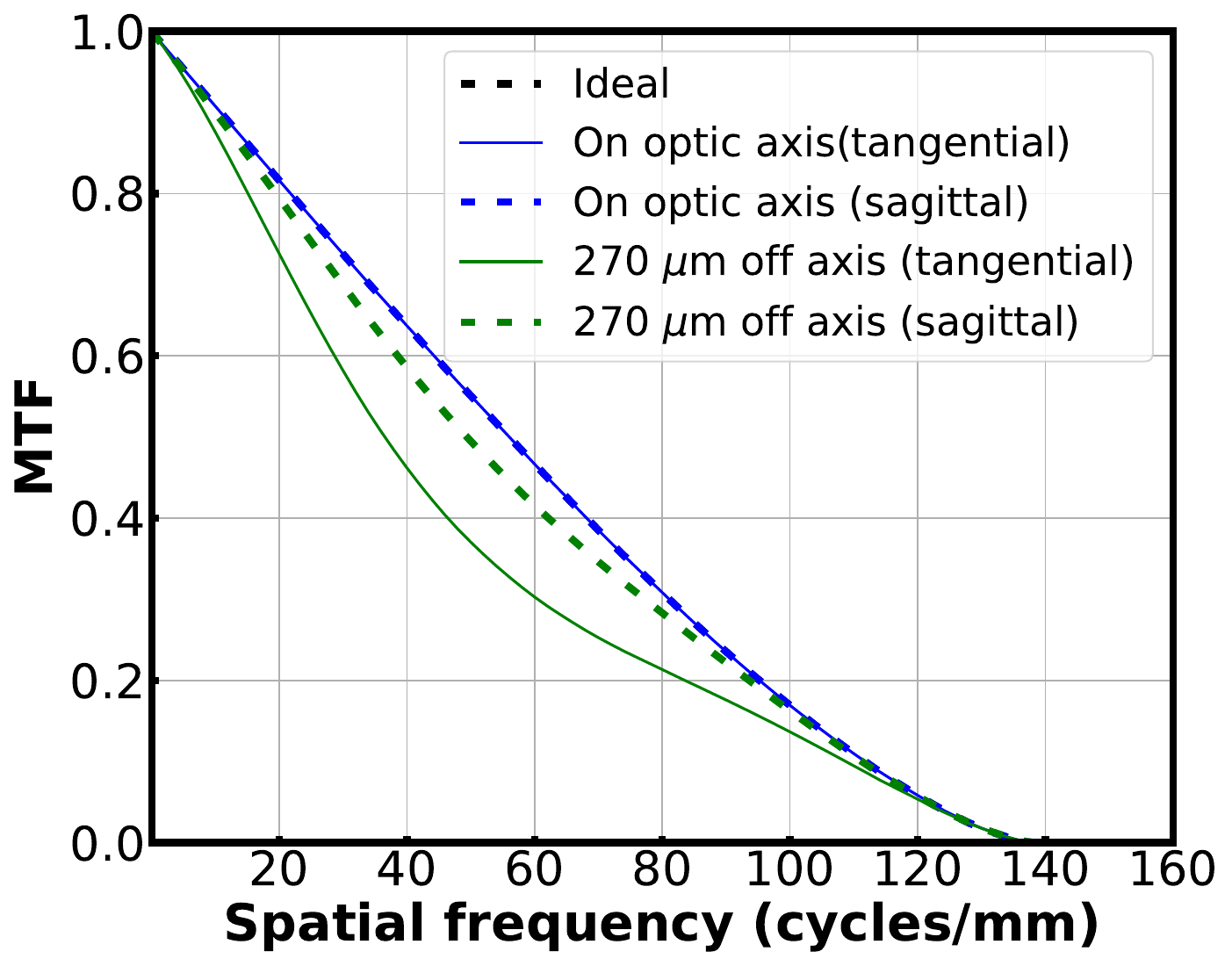}
   \end{subfigure}
    \caption{Simulation results for the example objective at 397~nm. (a) The calculated Strehl ratio (S) as a function of displacement from the optical axis. The dashed orange horizontal line indicates the diffraction-limited threshold ($\textrm{S} \ge 0.8$). The intersection point between S and the horizontal line occurs at radius of approximately 270~$\mu$m for a FOV of 540~$\mu$m. (b) The simulated MTF at the image plane for a point source on the optical axis and 270~$\mu$m off-axis. The ideal diffraction-limited MTF is also included for comparison (dashed black line) which largely overlaps with on-axis MTF curves. Note that Ansys Zemax OpticStudio optimizes the sagittal axis.
    \label{fig:simulation}
    }
\end{figure}

After finalizing the compound objective configuration with a target wavelength of 397~nm), simulation results show that the imaging system (including all viewports) offers (i) a working distance of 61~mm (from object to the first lens front surface), (ii) a back focal length of 613~mm (from the last lens to the image plane), (iii) a magnification of -10.6x, and (iv) on-axis diffraction-limited imaging performance with an RMS spot size smaller than the Airy disk radius of $r_\textrm{Airy} = 0.87~\mu\textrm{m}$. Given that experiments can involve imaging a long $^{40}$Ca$^+$ ion chain, we assess the system's off-axis performance, quantified by the FOV. Following the criterion outlined in Ref.~\cite{Kingslake2010LensDesign}, we simulate the Strehl ratio variation across the image focal plane, as shown in Figure~\ref{fig:strehl ratio}. In Ref.~\cite{Kingslake2010LensDesign} the FOV is defined as the area where the Strehl ratio remains $\textrm{S} \ge 0.8$. We simulate a FOV diameter of ${\sim}540~\mu\textrm{m}$, and compare the simulated modulation transfer function (MTF), both on-axis and at the edge of the FOV (270~$\mu$m off of the optical axis), to the diffraction-limited MTF shown in Fig.~\ref{fig:MTF}. We note a small deviation between the curves, suggesting that residual aberrations will not significantly degrade image contrast across this 540~$\mu$m diameter FOV.

\begin{table}[h!]
  \centering
  \caption{Parameters of example objective for other atomic neutral and ion species. The back working distance is constrained to 632.2~mm and the NA is restricted by the objective itself with a 50.8~mm diameter to 0.37. The distance between the outer cryostat window and the first compound objective lens ($d_\textrm{gap}$) is optimized for each listed wavelength. All other parameters are identical to the objective presented in Figure~\ref{fig:layout} with a design wavelength of 397~nm. The example objective remains diffraction-limited for all wavelengths listed.
  \label{tab:versatilely}
  }
  \begin{tabular}{lcccccl} 
\hline \hline
Species & Wavelength &$d_\textrm{gap}$ & NA & FOV &Magnification \\
 & (nm)& (mm)& (mm) & ($\mu$m) & (x)\\ 
\hline
Yb$^+$& 369 & 4.3 & 0.37 & 345 & -11.1\\
Ca$^+$ & 397 & 4.8 & 0.37 & 540 & -11.0\\
Sr$^+$& 422 & 5.2 & 0.36 & 608 & -10.9\\
Ba$^+$& 493 & 6.0 & 0.36 & 652 & -10.8\\
Rb&   780 & 7.4 & 0.35 & 720 & -10.5\\
Cs&   852 & 7.6 & 0.35 & 740 & -10.5\\
Sr$^+$ & 1092 & 8.1 & 0.35 & 800& -10.4\\
\hline

\end{tabular}
\end{table}

We optimize our example objective for monochromatic 397~nm light and do not correct for chromatic aberration. However, its performance at other wavelengths relevant to $^{40}$Ca$^{+}$ experiments is highly beneficial for laser beam alignment and other diagnostics. Simulating the sample objective design with a fixed back focal length while adjusting the gap distance (distance from the first lens to the closest window, noted as $d_\textrm{gap}$ in Fig.\ref{fig:layout}) for several relevant wavelengths (between 375~nm and 866~nm) shows that with only slight changes of $d_\textrm{gap}$ (-0.4~mm at 375~nm to +2.9~mm at 866~nm from the design wavelength of 397~nm) the design remains diffraction-limited. However, the magnification changes slightly, e.g. at 375~nm and 866~nm, the magnification changes to -10.7x and -10.1x, respectively. 

For ease of optical system assembly, maintaining high performance with small changes in the overall length of the imaging system is crucial. Our simulations show that the imaging system tolerates a 46~mm (-14~mm to +32~mm) variation in total length without compromising the diffraction-limited image quality for all $^{40}$Ca$^{+}$ relevant wavelengths. Note that a change in total length results in a change in magnification (-10.3 and -11.2, respectively) and FOV (690~$\mu$m and 230~$\mu$m, respectively). Critically, this change primarily affects the back focal length with minimal impact (${<}0.5$~mm) on $d_\textrm{gap}$ for a large range of target wavelengths. This relaxed tolerance requirement significantly simplifies the assembly process.

To demonstrate the versatility of this design, we simulate its performance at wavelengths relevant for other atomic neutral and ion species with results shown in Table \ref{tab:versatilely}. Note that we show the full NA without any experimental physical constraints, and we keep the total window thickness (13.4~mm, adding Veiwport, Window 1, and Window 2 in Table \ref{tab:prescription}). Our simulations indicate that maintaining diffraction-limited resolution with a few hundred $\mu$m FOV is achievable by simply adjusting $d_\textrm{gap}$ while maintaining a back working distance at 632.2~mm. For shorter or longer front working distances, the lens(es) and/or lens spacing(s) must be iteratively simulated and modified accordingly following the method described in Section~\ref{Sec:Method}.

\section{Diffraction limited performance}\label{Sec:Diffraction}

\begin{figure}[b!]
   \centering
   \begin{subfigure}[b]{0.48\textwidth}
     \centering   
     \caption{\label{fig:USAF aligned}}
      \includegraphics[width=\textwidth]{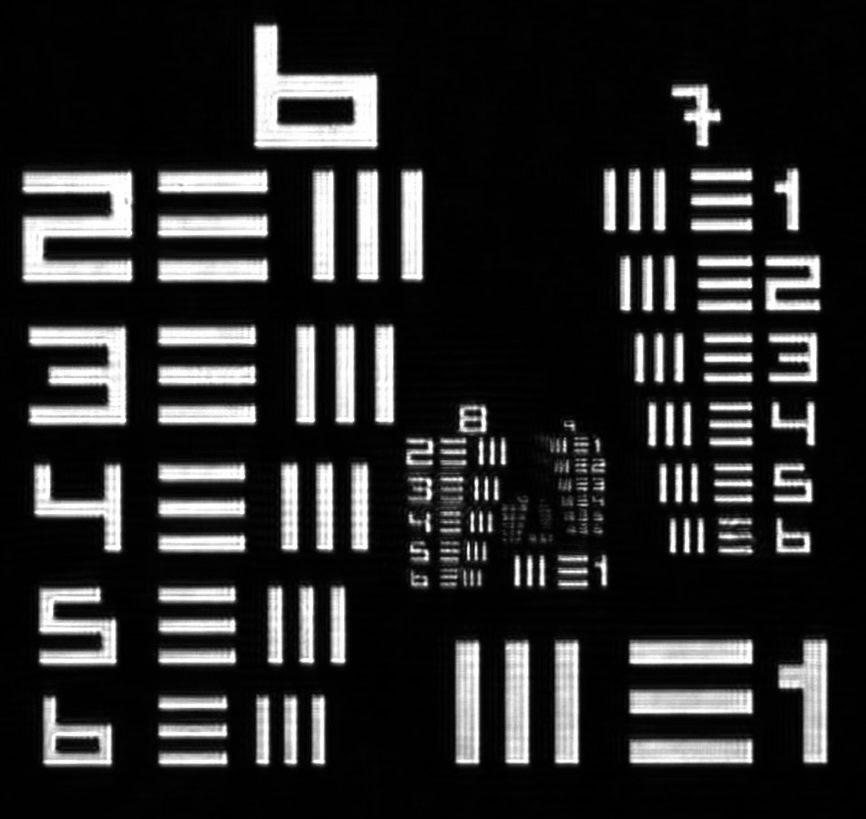}
   \end{subfigure}
   \begin{subfigure}[b]{0.455\textwidth}
     \centering
     \caption{\label{fig:USAF aligned zoom in}}
     \includegraphics[width=\textwidth]{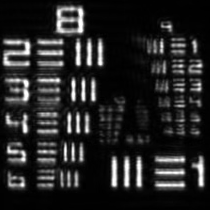}
   \end{subfigure}
    \caption{Image of a 1951 USAF resolution target under 397 nm diffused light illumination. (a) Full view, 2.99~mm wide in the imaging plane, showing resolved line pairs from Group 6 to Group 9. (b) Zoomed-in view, 0.72~mm wide, resolving Group 9 Element 2 (line width 0.87~$\mu$m).
    \label{fig:397 experiment}
    }
\end{figure}

After assembling the example objective described in Section~\ref{Sec:Design}, we validate the experimental performance. We diffuse and project 397~nm laser light on a 1951 USAF Resolution Test Target (Newport HIGHRES-1) through a series of optical flats and the assembled objective, using a camera (Thorlabs CS895MU) for image analysis. Note that all components are \textit{ex vacuo}. The optical flats are matched to the thickness and materials of our simulated vacuum and cryogenic viewport windows. An anodized aluminum aperture mask, sized to match the aperture-limited NA of 0.30 from our example apparatus, is positioned in front of the objective to block stray light. The distances between the test target, objective, and camera are controlled according to the parameters outlined in Table~\ref{tab:prescription}.

A full image of the resolution target, shown in Figure~\ref{fig:USAF aligned}, exhibits minimal distortions beyond diffraction-limit effects. A zoomed-in view of the same image is shown in Figure~\ref{fig:USAF aligned zoom in}. Here, the line pairs in Group 9, Element 2 are clearly resolved, corresponding to a resolution of ${\leq}0.87$~$\mu\textrm{m}$. Given the camera pixel size of $3.45 \times 3.45$~$\mu\textrm{m}^2$, we measure a system magnification of approximately $\left(-10 \pm 1\right)$x, which agrees with the -10.6x simulation result. However, small misalignments within the imaging system and diffraction fringes from the coherent 397 nm illumination result in some image quality degradation, as can be seen in Figure~\ref{fig:USAF aligned zoom in} with the presence of fringes.

\begin{figure}[t!]
   \centering
   \begin{subfigure}[b]{0.48\textwidth}
     \centering
     \caption{\label{fig:USAF 375}}
     \includegraphics[width=\textwidth]{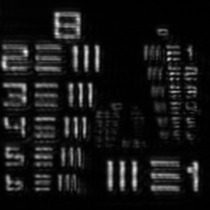}
   \end{subfigure}
   \begin{subfigure}[b]{0.48\textwidth}
     \centering
     \caption{\label{fig:USAF 866}}
     \includegraphics[width=\textwidth]{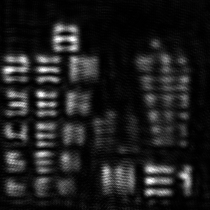}
   \end{subfigure}
   \caption{Images of a 1951 USAF resolution target under other wavelengths of interest for $^{40}$Ca$^+$. (a) 375 nm illumination, resolving Group 9 Element 2 (line width 0.87~$\mu$m) and partially resolving Element 3 (line width 0.75~$\mu$m), meeting the diffraction limit of 0.82~$\mu$m. (b) 866 nm illumination, resolving Group 8 Element 1 (line width 1.95~$\mu$m) and partially resolving Group 8 Elemenet 2 (line width 1.74~$\mu$m), meeting the diffraction limit of 1.90~$\mu$m.
   \label{fig:wavelengths}
    }
\end{figure}

\begin{figure}[b!]
   \centering
   \begin{subfigure}[b]{0.48\textwidth}
     \centering
     \caption{\label{fig:USAF 397 short}}
     \includegraphics[width=\textwidth]{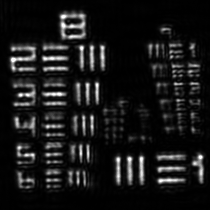}
   \end{subfigure}
     \begin{subfigure}[b]{0.48\textwidth}
     \centering   
     \caption{\label{fig:USAF 397 long}}
     \includegraphics[width=\textwidth]{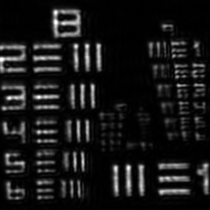}
   \end{subfigure}
    \caption{Images of a 1951 USAF resolution target under total length extremes under 397~nm illumination while maintaining diffraction-limited performance. (a) Total length (object to image) of approximately 714~mm. (b) Total length of approximately 760~mm. Both images show clear resolution of Group 9 Element 2, which meet the diffraction limit.
    \label{fig:versaltility}
    }
\end{figure}

We conduct further measurements to verify the sample objective's versatility across the full wavelength range of interest and to confirm its tolerance to variations in total length. Using the same resolution test target, we acquired images at the short- and long-wavelengths of interest (375~nm and 866~nm) with images shown in Figure~\ref{fig:wavelengths}. The measured $d_\textrm{gap}$ changes were consistent with our simulations, exhibiting a maximum deviation of 100 $\mu$m, which we attribute to combinations of multiple alignment stages' backlash and alignment precision. Additionally, we verified the example objective's tolerance to changes in total length at 397~nm. By adjusting the back focal length and $d_\textrm{gap}$, we demonstrate that the system maintains diffraction limited performance across 46~mm in total length variation as shown in Figure~\ref{fig:versaltility}.

\section{FOV measurement}\label{Sec:FOV}

Methods for characterizing the FOV have been explored in previous studies, yet each presents certain limitations. the approach in Ref.\cite{Bennie13Versatile} involved raster-scanning a pinhole across the object plane using automated motorized stages. The diffraction-limited FOV boundary was established when the image spot size remained below $\sqrt{2}$ times the on-axis spot size. Although this method generated a substantial dataset for analysis, it deviated from the widely accepted Strehl ratio S~$\leq0.8$ standard. Conversely, Ref.\cite{Li2020High} utilized the Strehl ratio as its reference for FOV definition but relied solely on on-axis and simulated FOV edges data, which lacked statistical rigor. To address these shortcomings, we present a novel method for quantitatively measuring the FOV by scanning a calibrated target across the object plane.

We characterize the example objective's FOV via a multi-step, sequential process.
\begin{enumerate}
    \item Precisely characterize both the imaging system magnification and the actual diameter of a 1~$\mu$m pinhole. 
    \item Simulate the ideal image generated by a pinhole with the measured actual diameter using the OpticStudio Physical Optical Propagation (POP) module as the ideal reference.
    \item Form a two-dimensional map of the Strehl ratio by comparing fitted camera images of the pinhole to the ideal reference; and
    \item Fit the resulting two-dimensional map to determine the FOV.
\end{enumerate}

\begin{figure}
\centering

\valign{#\cr
  \hbox{%
    \begin{subfigure}{.35\textwidth}
    \centering
    \caption{\label{fig:pinholeimage}}
    \includegraphics[width=\textwidth]{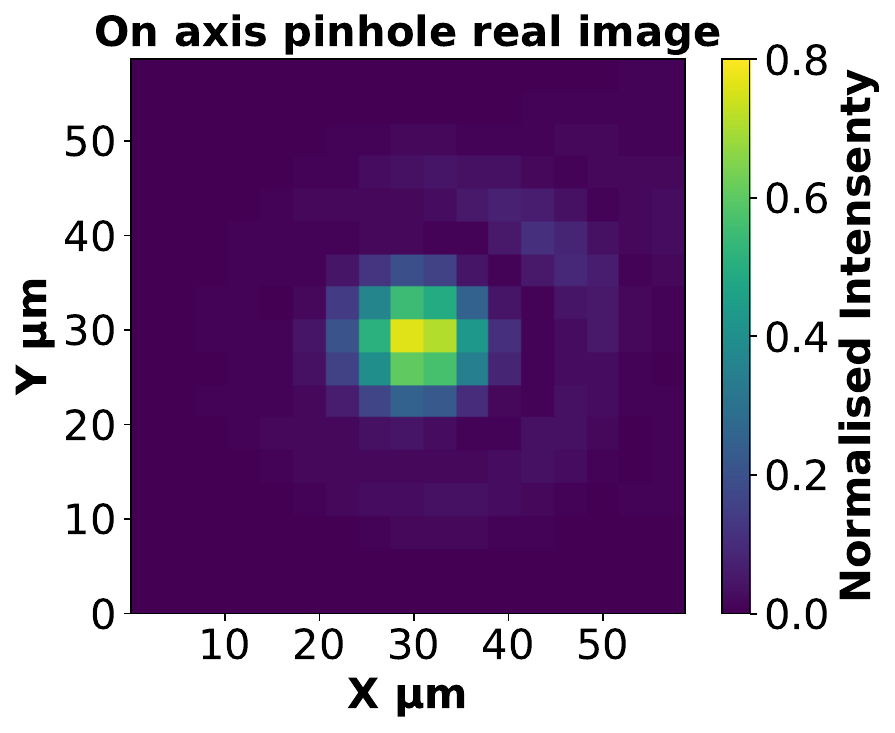}
    \end{subfigure}%
  }
  \hbox{%
    \begin{subfigure}{.35\textwidth}
    \centering
    \caption{\label{fig:088POP}}
    \includegraphics[width=\textwidth]{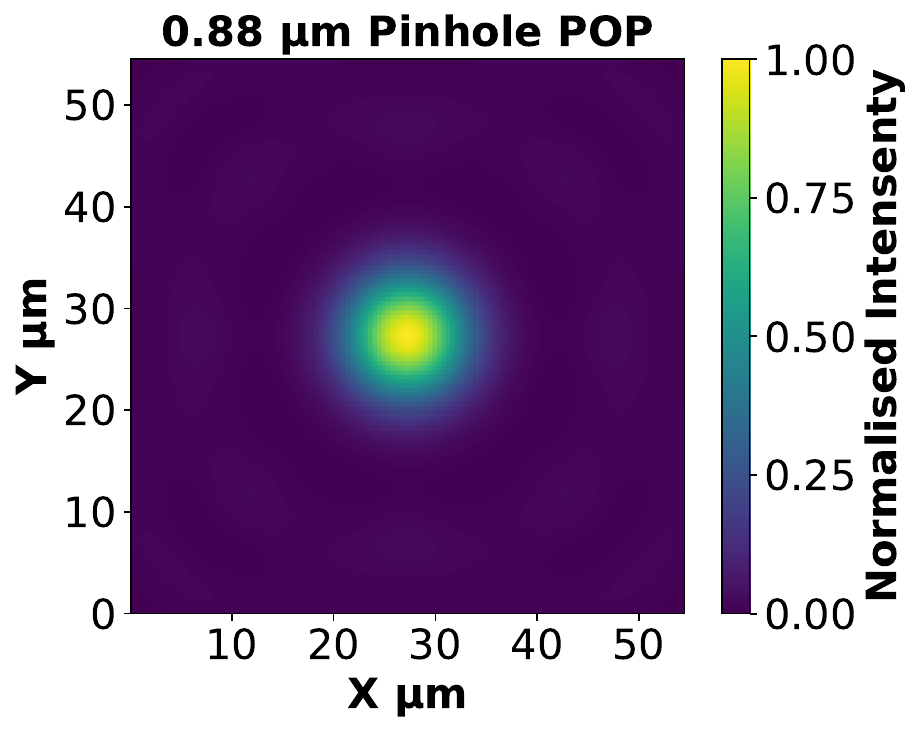}
    \end{subfigure}%
  }\cr
  \noalign{\hfill}
  \hbox{%
    \begin{subfigure}[b]{.64\textwidth}
    \centering
    \caption{\label{fig:FOV}}
    \includegraphics[width=\textwidth]{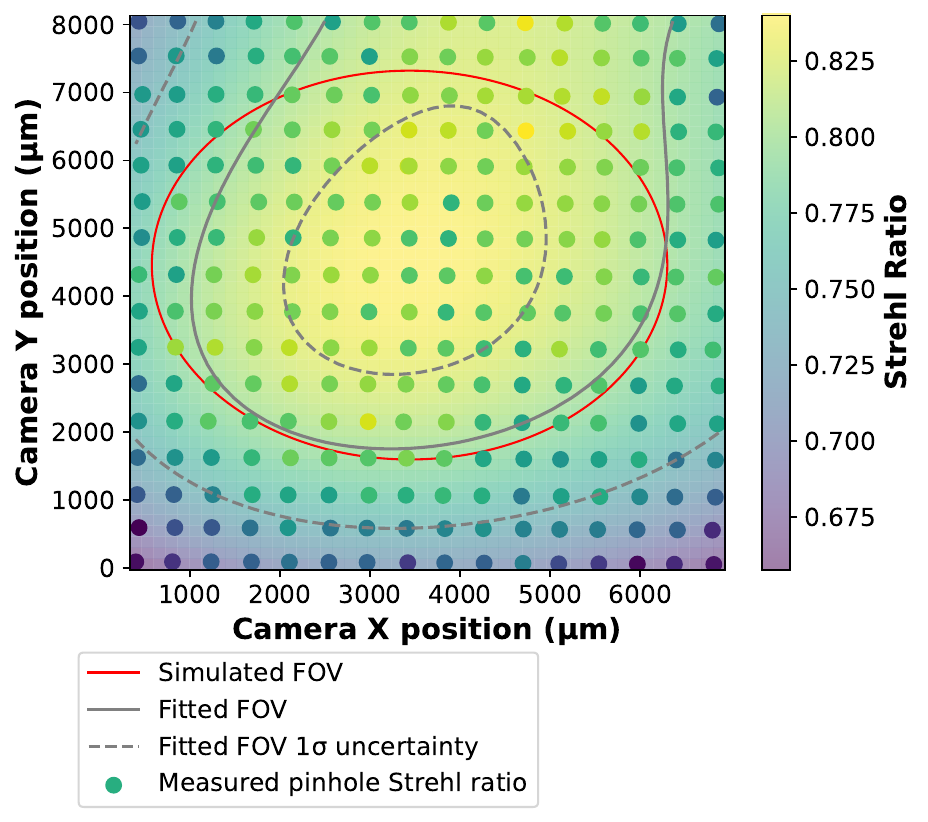}
    \end{subfigure}%
  }\cr
}

\caption{Experimental FOV results: (a) Image of Thorlabs P1K pinhole near the optical axis. (b) Ideal 0.88~$\mu$m pinhole image generated by POP. (c) Strehl ratio map across the camera sensor. The red solid contour is the 540~$\mu$m simulated FOV, the solid grey contour is the fitted FOV boundary to the experimental data, and the grey dashed lines are the fitted FOV $1\sigma$ uncertainty.}

\end{figure}

In Step~1, we first translate a 1~$\mu$m pinhole (Thorlabs P1K) across the object plane in a 4~$\times$~5~$\mu$m (horizontal~$\times$~vertical) grid pattern, acquiring 256~images on a camera (Thorlabs CC505MU). To maintain consistent illumination and precise positioning during data acquisition, we rigidly mount both the pinhole and the 397~nm light source to the same translation stage. Figure~\ref{fig:pinholeimage} shows a pinhole camera image near the optical axis center. By correlating the known displacement of the pinhole in the object plane with its corresponding movement on the camera, we determine that the magnification is -10.6(4)x with the uncertainty attributed to \sout{micrometer's} stage backlash.

In Step~2, we generate an ideal reference image for the real pinhole. The ideal pinhole image is the convolution of the diffraction-limited Airy disk at 397 nm of the objective and the -10.6(4)x magnification of the actual pinhole. Using this as a fitting model, and measuring distance from the bright peak value to the first dark ring of Figure~\ref{fig:pinholeimage}, we deduce that the actual pinhole size is 0.88(3)~$\mu$m, which agrees with the pinhole manufacturing tolerances. We then use the OpticStudio Physical Optical Propagation module to generate the ideal reference image of the light passing through the 0.88~$\mu$m pinhole.

In Step~3, we determine the Strehl ratio by comparing all 256 experimental pinhole images with the simulated ideal reference. We first normalize the total light power of each experimental image to that of the ideal image (Figure~\ref{fig:088POP}. We then calculate the Strehl ratio as the ratio of the experimental fitted peak value to that of the ideal image. We iterate this calculation for all 256 pinhole images. The resulting Strehl ratios are plotted as a function of pinhole coordinates, generating the 16~$\times$~16 map shown in Figure~\ref{fig:FOV}.

In Step~4, we provide a visual representation of the Strehl ratio distribution by fitting a third-order polynomial to the experimental Strehl ratio values to generate a two-dimensional map as the background to the 16$\times$16 experimental points, shown in Figure~\ref{fig:FOV}. We determine the experimental field of view (FOV) boundary as the contour line corresponding to a Strehl ratio of 0.8 (solid grey line in Figure \ref{fig:FOV}, with 1$\sigma$ confidence interval indicated by dashed grey lines. The FOV boundary from simulation (red circle in Figure~\ref{fig:FOV}) is consistent with the simulation results in Section~\ref{Sec:Design}, confirming the validity of our approach.

\section{Conclusion}\label{Sec:Conclusion}
We present a robust and versatile objective design utilizing readily available COTS components to achieve diffraction-limited performance with a long front working distance and high NA. The design employs a compound lens configuration, optimized to correct for spherical aberrations introduced by optical windows between the object and the objective. Additionally, the objective directly converges captured light to the back focal plane, simplifying construction and alignment. Our example objective presented in Section~\ref{Sec:Design}, designed for 397~nm light using the methodology described in Section~\ref{Sec:Method}, experimentally demonstrates a long front working distance of 61~mm, a clipped NA of 0.30, a resolution of 0.87~$\mu$m, and a a 540~$\mu$m FOV, which all agree with simulation. Further experimental results confirm diffraction-limited resolution from 375~nm to 866~nm with the same objective lens, and its performance remains robust across a 46 mm range of total length adjustment. In addition to the objective design itself, we also develop a quantitative method to measure the field of view (FOV) using an experimentally-calibrated pinhole target. This method provides a valuable tool for accurately characterizing the imaging performance of optical systems and complements traditional resolution measurements. The combination of our design methodology's high optical performance, mechanical robustness, and practical advantages makes this objective lens design a valuable asset for a wide range of AMO physics and QIS experiments, facilitating the exploration of diverse quantum phenomena.

\section{Acknowledgments}\label{Sec:Acknowledgments}
Portions of this research was sponsored by the Laboratory Directed Research and Development Program of Oak Ridge National Laboratory, which is managed by UT-Battelle, LLC, for the US Department of Energy. Additional portions of this material is based upon work supported by the U.S. Department of Energy, Office of Science, National Quantum Information Science Research Centers, Quantum Science Center. We would like to thank John Huggins and Chelo Chavez for their contributions in engineering the spacer rings.The authors declare no conflicts of interest.


\bibliography{Main}

\end{document}